\documentclass[aps,pra,reprint,a4paper,superscriptaddress,numbers,longbibliography]{revtex4-1}

\usepackage[pdftex]{hyperref,color,graphicx}
\usepackage{amsfonts,amssymb,amsmath}
\usepackage{enumitem}
\usepackage[separate-uncertainty=true]{siunitx}
\usepackage[english]{babel}
\usepackage[utf8]{inputenc}
\usepackage[T1]{fontenc}

\hypersetup{
	pdftitle = {Atomic "bomb testing": the Elitzur-Vaidman experiment violates the Leggett-Garg inequality},
	pdfauthor = {Carsten Robens, Wolfgang, Alt, Clive Emary, Dieter Meschede, Andrea Alberti},
	colorlinks=true,linkcolor=blue,citecolor=blue,filecolor=blue,urlcolor=blue
}

\newif\ifusebibfile
\usebibfilefalse

\newcommand{\figref}[2]{\hyperref[#1]{\ref{#1}(#2)}}
\newcommand{\ket}[1]{|{#1}\rangle}
\newcommand{\definition}{{\lower-0.3pt\hbox{:}\kern-3pt=}}
\renewcommand\textemdash{\leavevmode\unskip\kern0.8pt\rule[0.215\baselineskip]{8pt}{0.22pt}\kern1pt\ignorespaces}

\makeatletter
\def\frontmatter@abstractwidth{\textwidth}%
\def\subsection{%
  \@startsection
    {subsection}%
    {2}%
    {\z@}%
    {.8cm \@plus1ex \@minus .2ex}%
    {.5cm}%
    {%
     \normalfont\small\bfseries
	 \centering
    }%
}%
\makeatother

\renewcommand{\subsubsection}[1]{{\normalfont\itshape{}#1.}}

\begin{document}
\selectlanguage{english}

\title
{
Atomic \texorpdfstring{``bomb testing''\hspace{0.3pt}}{"bomb testing"}: the Elitzur-Vaidman experiment violates the Leggett-Garg inequality\\[3mm]
}

\author{Carsten Robens}
\author{Wolfgang Alt}
\affiliation{Institut für Angewandte Physik, Universität Bonn,
Wegelerstr.~8, D-53115 Bonn, Germany}
\author{Clive Emary}
\affiliation{Joint Quantum Centre Durham-Newcastle, Newcastle University, Newcastle upon Tyne NE1 7RU, UK}
\author{Dieter Meschede}
\affiliation{Institut für Angewandte Physik, Universität Bonn,
Wegelerstr.~8, D-53115 Bonn, Germany}
\author{Andrea Alberti}
\email{alberti@iap.uni-bonn.de}
\affiliation{Institut für Angewandte Physik, Universität Bonn,
Wegelerstr.~8, D-53115 Bonn, Germany}
\date{\today}
\begin{abstract}
{\bfseries{Abstract}.
Elitzur and Vaidman have proposed a measurement scheme that, based on the quantum superposition principle, allows one to detect the presence of an object \textemdash in a dramatic scenario, a bomb \textemdash without interacting with it.
It was pointed out by Ghirardi that this interaction-free measurement scheme can be put in direct relation with falsification tests of the macro-realistic worldview.
Here we have implemented the ``bomb test'' with a single atom trapped in a spin-dependent optical lattice to show explicitly a violation of the Leggett-Garg inequality \textemdash a quantitative criterion fulfilled by macro-realistic physical theories.
To perform interaction-free measurements, we have implemented a novel measurement method that correlates spin and position of the atom.
This method, which quantum mechanically entangles spin and position, finds general application for spin measurements, thereby avoiding the shortcomings inherent in the widely used push-out technique.
Allowing decoherence to dominate the evolution of our system causes a transition from quantum to classical behavior in fulfillment of the Leggett-Garg inequality.
}
\end{abstract}

\maketitle
\section{Introduction}

Measuring physical properties of an object \textemdash whether macroscopic or microscopic \textemdash is in most cases associated with an interaction.
For example, scattering photons off an object allows one to detect its presence in a given region of space.
However, this also produces a small perturbation of its state by direct momentum transfer.
It is well known from numerous discussions on the physics of the quantum measurement process (see, e.g., Refs.~\cite{Wheeler:1983, Leggett:2005}) that a measurement in general modifies the quantum evolution unless the object is already in an eigenstate of the measurement apparatus \cite{Lueders:1951}.
This is even the case when the measurement yields a negative outcome, that is, when we did \emph{not} find the particle on a certain trajectory that had originally a non-vanishing probability amplitude to be occupied.
For example, in a double-slit experiment, quantum interference is suppressed as soon as a measurement detects the which-way information, regardless of whether the information is acquired by direct interaction or indirect negative inference.
Quantum mechanics formalizes the loss of interference in terms of the quantum measurement process, showing that measurements are generally invasive as they entail a modification of the subsequent quantum evolution.
While the quantum measurement process is still intensely debated in the literature \cite{Bassi:2013}, we adopt here the pragmatic view that a measurement applied to a superposition state causes a sudden reduction of the wave function to a smaller Hilbert space.

Ideal negative measurements, that is, measurements without direct interaction, play an important role in a physical scenario known as \emph{macro-realism} \cite{Ghirardi:1986,Diosi:1989,Pikovski:2015}.
This scenario advocates a classical worldview describing the state of macroscopic objects, according to which macroscopic objects are always in one of several possible macroscopically distinct states.
In a macro-realistic worldview, it is plausible to assume that a negative outcome of a measurement cannot affect the evolution of a macroscopic system, meaning that negative measurements are \emph{non-invasive} \cite{Emary:2013}.
In order to rigorously test the macro-realistic point of view, Leggett and Garg have derived an inequality from the assumptions of macro-realism and non-invasive measurability, which can be violated by quantum-mechanical superposition states \cite{Leggett:1985}.
Relying on the implementation of an ideal negative-measurement protocol \textemdash a prerequisite for any genuine test of the Leggett-Garg inequality \textemdash experimental violations of the macro-realistic worldview have been experimentally demonstrated with phosphor impurities in silicon by Knee et al.~\cite{Briggs:2012} and with trapped atoms by Robens et al.~\cite{Robens:2015}.

The definition of the degree of ``macroscopic distinctness'' has been a matter of discussion in the literature ever since \cite{Leggett:2002}, and is likely to remain as such till an experiment \cite{Arndt:2014} will shed new light, for example, discovering a physical ``classicalization'' mechanism that causes an objective reduction of wave packets.
Recently, Nimmrichter and Hornberger proposed a quantitative criterion based on a minimal macro-realistic extension of quantum mechanics to quantify the macroscopicity of an object \cite{Nimmrichter:2013}. 
Their objective criterion of macroscopicity allows us to experimentally test the behavior of a single trapped atom \textemdash however microscopic it is, according to our intuition \textemdash under the hypothesis of macro-realism, as we can put its macroscopicity directly in relation to that of other, ideally more massive physical objects.

It was pointed out by Ghirardi \cite{Ghirardi:2007} that a Leggett-Garg test of macro-realism is naturally related to the notion of interaction-free measurements introduced by Elitzur and Vaidman \cite{EV:1993}.
In a rather dramatic scenario, Elitzur and Vaidman proposed a quantum device able to single out live ``bombs'' from a collection containing also duds without triggering them nor interacting with them.
While the first realizations of the Elitzur-Vaidman experiment employed ``flying'' photons \cite{Kwiat:1995} and ``flying'' neutrons \cite{Hafner:1997}, we here implement a variation of this experiment with neutral atoms trapped in a one-dimensional optical lattice.
A convenient scheme for interaction-free measurements with neutral atoms has been demonstrated by Robens et al.~\cite{Robens:2015} exploiting state-dependent optical potentials.
Following the idea of Ghirardi, we tested the hypothesis of macro-realism with our atomic implementation of the Elitzur-Vaidman ``bomb testing'' \emph{Gedankenexperiment}.
Our experiment shows explicitly that the Leggett-Garg inequality is violated by 21\,$\sigma$.
In addition, trapped atoms can be held for long times.
By controlling the duration of a suitably chosen wait interval, it is straightforward to study the influence of decoherence and experimentally observe the gradual transition from quantum to classical behavior.

It is our pleasure to honor with these recent experimental results Theodor W.\ Hänsch.
For many decades he has laid the foundations in laser physics and technology, without which present-day laser control of quantum particles is hardly conceivable.
In this sense, the objective of this article is to experimentally demonstrate yet another example of exploring the world with lasers and laser-controlled atoms at the quantum-classical boundary.

\section{Brief review: the Elitzur-Vaidman interaction-free \texorpdfstring{\textquotedblleft}{"}bomb test\texorpdfstring{\textquotedblright}{"}}
\label{sec:EVbombTest}

Let us illustrate the concept of the interaction-free ``bomb test'' as presented in the original work by Elitzur and Vaidman \cite{EV:1993}, based on a single photon travelling on a quantum superposition of trajectories along the two paths of a Mach-Zehnder (MZ) interferometer.

\begin{figure}[b]
	\centering
	\includegraphics[width=\columnwidth]{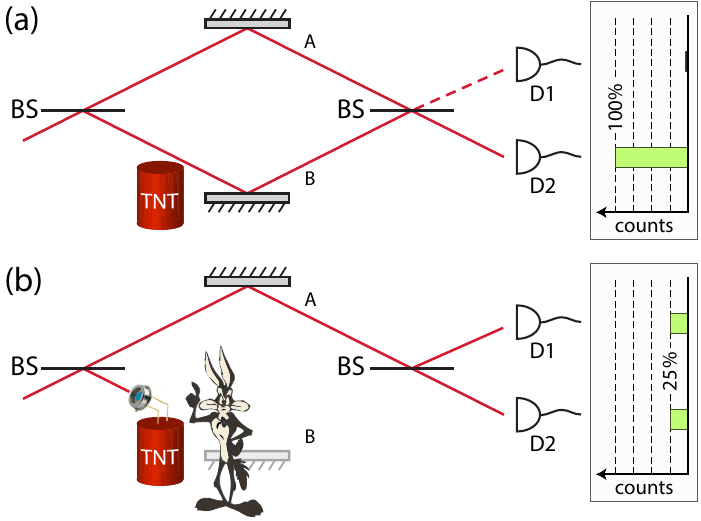}
	\caption{
	Bomb testing with a Mach-Zehnder interferometer operated with a single
    photon at a time.
	The beam splitters have split ratio 50:50.
	(a) The phase difference between the two arms is adjusted such that all photons are directed to detector D2.
	The object situated close to the lower branch B (e.g., a dud bomb not equipped with the trigger) does not intercept the photons.
	(b) The object (e.g., a ``live bomb'' equipped with an avalanche photodiode which is triggered by a single photon \cite{Wile}) intercepts the photons in the lower branch B.}
	\label{fig:MZI}
\end{figure}

The two branches A and B of a MZ interferometer can be balanced in such a way that one of the two output ports, say D2, is always bright while the other one, D1, is always dark, see Fig.~\figref{fig:MZI}{a}.
If any object (e.g, a ``live bomb'' triggered by a single photon) intercepts the trajectory of the photon in the lower branch B, the photon field is removed from that branch and the balance of the MZ interferometer is disturbed.
As shown in Fig.~\figref{fig:MZI}{b}, the photon has a $\SI{50}{\percent}$ probability to be  blocked without ever reaching the detectors.
This event is of course highly invasive: the ``bomb'' is triggered to explode by absorbing the travelling photon.

In all other events, the photon must have followed branch A, thus avoiding interacting with the object, and is subsequently routed to the detectors D1 and D2 with equal probability.
In case detector D2 clicks, insufficient knowledge is gained to conclude on whether the object is present, as this outcome also occurs for no object present (``dud bomb'').
If the dark detector D1 lights up, however, the presence of an object in one of the two arms is signaled with certainty.
Since the photon could not reach the detector if it had touched the object, finding a photon with detector D1 detects the presence of the object without touching it, therefore leaving the ``bomb'' intact.
Following Elitzur and Vaidman, we call this measurement scheme \emph{interaction-free} \cite{Vaidman:2003}.

We obtain further insight in the Elitzur-Vaidman ``bomb test'' by analyzing its outcomes from the perspective of statistical hypothesis testing, where the Elitzur-Vaidman quantum device is employed to test whether the ``bomb'' is live (\emph{positive} test result).
The typical figure of merit in hypothesis testing is the statistical power $1-\beta$, that is, 
the fraction of ``live bombs'' which are correctly identified (without being triggered) and rescued.
In their original work \cite{EV:1993}, Elitzur and Vaidman showed that this fraction amounts to \SI{25}{\percent} (\emph{true positives}).
It is worth extending the analysis of ``bomb testing'' to quantum devices that, under realistic conditions, are impaired by decoherence.
We assume that decoherence reduces the fringe amplitude of the MZ interferometer in an unbiased way, resulting in a contrast $C$ less than unity.
In the presence of decoherence, the statistical power of the test remains unchanged, since this quantity describes the situation of a ``live bomb'', where the coherence of the photon plays no role, see Fig.~\figref{fig:MZI}{b}.
We therefore introduce a second figure of merit accounting for decoherence, which is given by the statistical error of type I, that is, the probability $\alpha$ that we erroneously rescue the bomb even though the bomb is a dud (\emph{false positive}).
One can show that the probability of type I errors is $\alpha=(1-C)/2$.
This measure vanishes only for a decoherence-free ``bomb tester'' and becomes $\alpha=1/2$ in the completely incoherent limit (equivalent to random selection).
It is also worth mentioning that, by allowing for repeated measurements (this is always possible until the ``bomb'' has not exploded), the statistical power $1-\beta$ can be straightforwardly increased to $1/3$ for 50:50 beam splitters and to $1/2$ for beam splitters with different split ratios.
Furthermore, a more complex variant of this scheme based on the quantum Zeno effect has been suggested \cite{Kwiat:1994} and experimentally demonstrated using traveling photons \cite{Kwiat:1999}, allowing one to approach unity efficiency ($\beta=0$).

Let us now reconsider the situation illustrated in Fig.~\figref{fig:MZI}{b} from the perspective of a macro-realist, who conducts the same experiment but with a massive, ideally macroscopic object traveling along the two branches of the MZ interferometer.
For a macro-realist, the massive particle travels either along branch A or along branch B, but not in a superposition state of the trajectories A and B.
If one discards by post-selection all events where the particle's presence on one of the two trajectories has been directly detected through the absorption by the object, then only interaction-free measurements (i.e.~ideal negative measurements) of the particle's position are considered.
By intercepting the particle at one time in branch A and at another time in branch B, the macro-realist learns about the particle's position in the MZ interferometer avoiding any interaction with it.
Therefore, to a macro-realist, ideal negative measurements must appear \emph{non-invasive}, since the subsequent evolution of the particle could have not been influenced by the presence of an object where the particle was \emph{not}.
Non-invasive measurements constitute an important prerequisite for any rigorous test of macro-realism through a violation of the Leggett-Garg inequality \cite{Emary2013a}.

\section{Relation to the Leggett-Garg inequality}
\label{sec:LeggettGarg}
Based on two assumptions, (A1) macro-realism and (A2) non-invasive measurements,
the Leggett-Garg inequality bounds a linear combinations of two-time correlation measurements,
\begin{equation}\label{eq:LG}
    K \hspace{-0.25pt}=\hspace{-0.25pt} \langle Q(t_2)Q(t_1)\rangle\hspace{-0.4pt}+\hspace{-0.4pt} \langle Q(t_3)Q(t_2)\rangle \hspace{-0.4pt}-\hspace{-0.4pt} \langle Q(t_3)Q(t_1)\rangle \leq 1,
\end{equation}
where $Q(t_i)$ denote the outcome of measurements carried out on the object at three subsequent times ($i=1,2,3$).
The values assigned to the individual measurement outcomes have to fulfill the condition $|Q(t_i)|\leq 1$ but can otherwise be freely chosen.
Standard quantum mechanics and macro-realism are distinguished from each other by violating and fulfilling this inequality, respectively.

Observing a violation of the inequality (\ref{eq:LG}) allows one to refute (i.e., \emph{falsify}) the assumptions underlying the Leggett-Garg test in the range of parameters investigated by the experiment, which in general are represented by the mass of the superposition states and their spatial split distance \cite{Nimmrichter:2013}.
For simple logical reasons, a violation of Eq.~(\ref{eq:LG}) implies that \emph{at least one} of the two assumptions (A1) and (A2) must not hold.
In conducting a Leggett-Garg test, it is therefore crucial that assumption (A2) of non-invasive measurability cannot be simply dismissed by a macro-realist claiming that a measurement operation, due experimental inadvertence, has influenced the subsequent evolution of the particle;
or else, even in case of an observed violation, no claim can be made concerning assumption (A1) of macro-realism.
In the literature, this type of objection is addressed as the \emph{clumsiness} loophole \cite{Wilde:2012}.
To circumvent it, Leggett and Garg suggested using ideal negative measurements, which, to say it à la Vaidman and Elitzur, are interaction free.
It should also be noted that in standard quantum theory both assumptions do not hold.
In fact, standard quantum theory postulates (1) no limit on the mass and split distance of a superposition state and that (2) even a negative measurement can cause the wave packet's reduction.
The latter point is indeed central to the Elitzur-Vaidman interaction-free experiment.

In the Elitzur-Vaidman experiment, the object that is put to the test of the Leggett-Garg inequality is the single photon travelling along the two paths of the MZ interferometer.
In the following we define the three measurement operations performed on the photon and their assigned values $Q(t_i)$, which are employed to perform the Leggett-Garg test in Eq.~\eqref{eq:LG}:
\begin{enumerate}[leftmargin=11.2mm,label=$Q(t_{\arabic*})$:]
	\item{}We identify the first measurement at $t_1$ with the preparation of the initial state \textemdash a photon in the input of the MZ interferometer.
	This measurement is by definition non-invasive, as it leaves the particle in the initial state.
	We assign to this measurement the value $Q(t_1){=}{+}1$.
	\item{}This measurement is performed at time $t_2$ when the photon is either in branch A or B of the MZ interferometer.
	It detects in which branch the photon travels by removing the photon at one time from branch A, at another time from branch B.
	The events in which the photon was directly intercepted by the object (i.e., when the ``bomb'' exploded) are discarded by post-selection in order to ensure that only ideal negative measurements are performed;
	this measurement must thus appear to a macro-realist as non-invasive  as it avoids any direct interaction with the photon itself.
	We assign to this measurement the constant value $Q(t_2){=}{+}1$ regardless of which trajectory the photon has taken.
	It should be noted that this measurement is not performed when evaluating the correlation function $\langle Q(t_3)Q(t_1)\rangle$ of the Leggett-Garg test.
	In fact, from the perspective of a macro-realistic who advocates (A1) and (A2), an ideal negative measurement could not have not influenced the evolution of the particle.
	\item{}The final measurement is performed at time $t_3$ when the photon has reached the output of the MZ interferometer.
	Depending on whether detector D1 or D2 has clicked, we assign to this measurement the value $Q(t_3){=}{+}1$ or $Q(t_3){=}{-}1$, respectively.
	Because we are not interested in the system's evolution after $t_3$, non-invasiveness of this measurement operation is not required.
\end{enumerate}
Previous Leggett-Garg experimental tests prior to Robens et al.~\cite{Robens:2015} only considered dichotomic designations of $Q(t_2)$, as opposed to the constant choice of $Q(t_2){=}{+}1$ here.
By deliberately disregarding the unnecessary, dichotomic constraint, we allow the Leggett-Garg correlation function of the Elitzur-Vaidman experiment to reach the maximum value, $K=2$, as permitted by quantum theory for a two level system (i.e., the photon in a superposition state of the trajectories A and B).
It can be shown that a violation of the inequality in Eq.~\eqref{eq:LG} is also produced in the Elitzur-Vaidman experiment for a dichotomic choice of $Q(t_2)$ (see Appendix~\ref{app:dichotomic}).
However, in this case the maximum value of $K$ predicted by quantum theory is only $3/2$ \cite{Budroni:2013,Schild:2015} instead of $2$.

Taking into account our specific designation of $Q(t_i)$, we can recast Eq.~\eqref{eq:LG} into a simpler form.
Since $Q(t_2){=}1$, the correlation function $\langle Q(t_3)Q(t_2)\rangle$ equals $\langle Q(t_3)\rangle_{\text{with~}Q2}$, that is, the average value of $Q(t_3)$ conditioned on a negative result of the measurement $Q(t_2)$.
Likewise, since $Q(t_1){=}1$, the correlation function $\langle Q(t_3)Q(t_1)\rangle$ simplifies to $\langle Q(t_3)\rangle_{\text{without~}Q2}$, that is, the average value of $Q(t_3)$ without having measured the position of the photon at $t_2$.
Our experiment can thus be analyzed with a simplified version of Eq.~\eqref{eq:LG},
\begin{equation}
	\label{eq:LGsimp}
	K = 1 + \langle Q(t_3)\rangle_{\text{with~}Q2} - \langle Q(t_3)\rangle_{\text{without~}Q2} \leq 1 \,.
\end{equation}

It is interesting to go one step further to provide an interpretation of the Leggett-Garg correlation function $K$ from the point of view of quantum theory.
We know that the values of $Q(t_3){=}\pm1$ are evenly distributed when the $Q(t_2)$ measurement is performed, since the latter reveals the which-way information; hence $\langle Q(t_3)\rangle_{\text{with~}Q2}\hspace{2pt}{=}\hspace{2pt} 0$.
Moreover, one can prove that $\langle Q(t_3)\rangle_{\text{without~}Q2}$ is identical to the contrast $C$ of the MZ interferometer (see Appendix~\ref{app:RamseyContrast}).
Thus, we find that the correlation function $K$ takes the suggestive form \begin{equation}
\label{eq:LGRamsey}
K = 1 + C\,.
\end{equation}
Intuitively, the function $K$ provides a quantitative indication, say a witness, of the amount of superposition involved in the evolution of the quantum particle.
Note that $K$ can be put in relation to the first quantum witness $W$ of superposition states introduced in Ref.~\citenum{Che-Ming:2012} (also described as {no-signaling in time} in Ref.~\citenum{Brukner:2013}).
In fact, $W=|K-1|=C$ as shown in Ref.~\citenum{Robens:2015}.
This demonstrates that the figure of merit $\alpha$ of a partially decohered ``bomb tester'' (see Sec.~\ref{sec:EVbombTest}) is directly related to the quantum witness $W$ since $\alpha=(1-W)/2$.
Furthermore, our results shows that any quantum particle exhibiting a non-vanishing interference contrast should allow, according to quantum theory, for a violation of Eq.~\eqref{eq:LG}, provided that one can additionally show through an experiment that the which-way information acquired through interaction-free measurements yields a vanishing contrast, that is, $\langle Q(t_3)\rangle_{\text{with~}Q2}\hspace{2pt}{=}\hspace{2pt} 0$.

\begin{figure*}[t]
	\centering
	\includegraphics[width=0.9\textwidth]{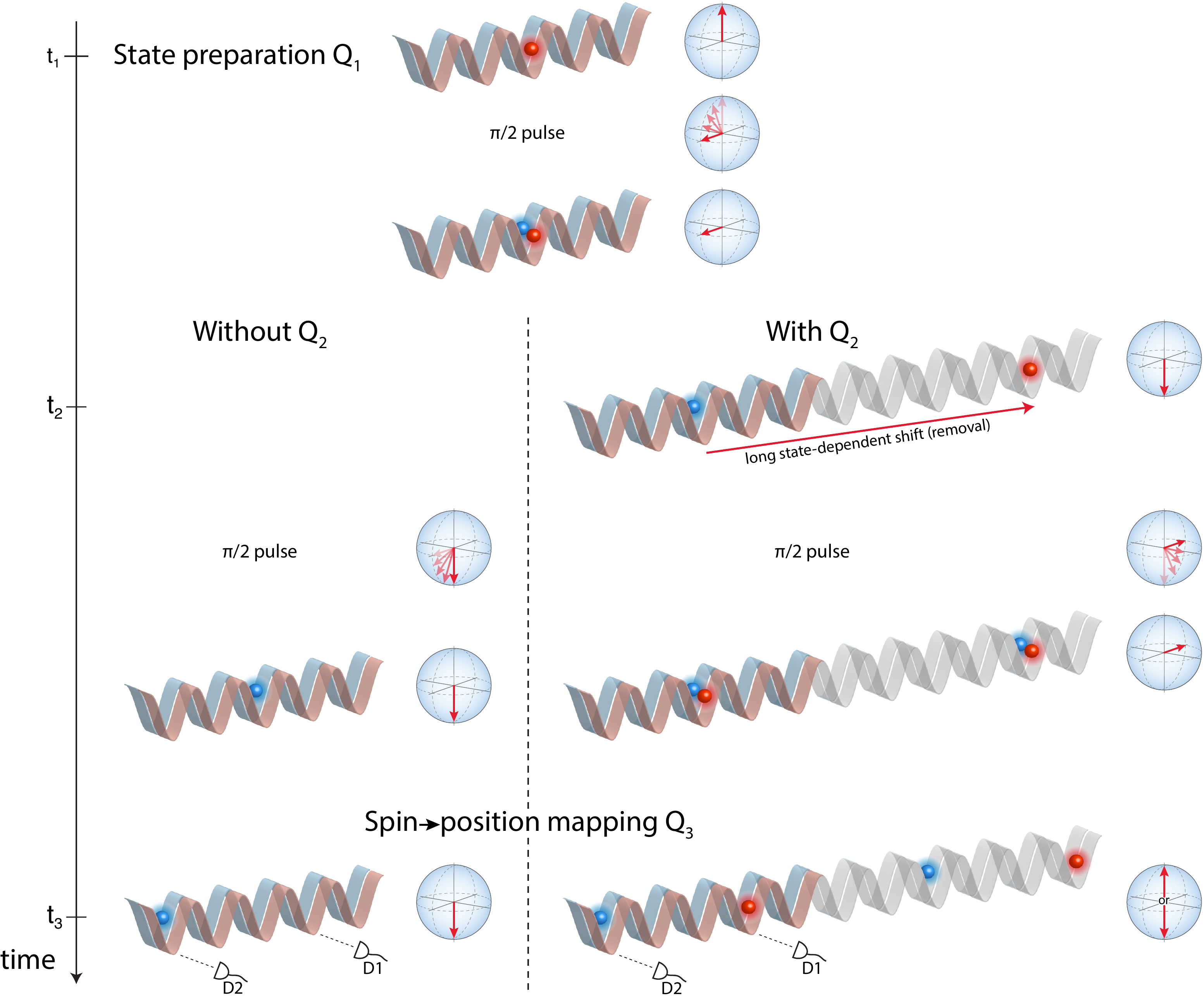}
	\caption{Illustration of the Elitzur-Vaidman experiment using single atoms.
	Atoms are trapped in state-dependent optical lattices, which consist of two independently movable, periodic optical potentials for atoms in the internal states $\ket{\uparrow}$ and $\ket{\downarrow}$.
The two atomic states form a spin-1/2 system, which is represented on the Bloch sphere at different moments of the time evolution; short microwave pulses allow us to rotate the spin.
On the left-hand side, protocol of a Ramsey interferometer, whose pulses are configured to produce the state $\ket{\downarrow}$;
this situation is equivalent to that in Fig.~\figref{fig:MZI}{a}.
The spin information is eventually mapped onto two different positions on the lattice, D1 and D2, which are efficiently detected by fluorescence imaging.
On the right-hand side, protocol of a Ramsey interferometer where an interaction-free measurement (i.e., an ideal negative measurement) of the spin state is performed at the intermediate time $t_2$.
This measurement intercepts only atoms in one spin state by transporting them far apart (grayed lattice regions); this situation is equivalent to that in Fig.~\figref{fig:MZI}{b}.
}
	\label{fig:MZIa}
\end{figure*}

\section{Interaction-free measurement with trapped atoms}
\label{sec:EV_exp}

Our atomic realization of the Elitzur-Vaidman experiment employs single neutral atoms trapped in an optical potential instead of flying photons.
Instead of delocalizing the particle on two distinct trajectories as in the MZ interferometer shown in Fig.~\ref{fig:MZI}, we let the particle evolve in a superposition of two long-lived internal states, which we denote by $\ket{\uparrow}$ and $\ket{\downarrow}$ hereafter.
In our experiment, an atomic Ramsey interferometer plays the role of the optical MZ interferometer.

\subsection{Experimental apparatus}
\label{sec:ExpApparatus}
\subsubsection{State-dependent optical conveyor belts}
At the core of our realization of the Elitzur-Vaidman experiment with trapped atoms are polarization-synthesized (PS) optical lattices, which were recently introduced by Robens et al.~\cite{Robens:2016}:
two one-dimensional, periodic optical potentials can be independently shifted along their common longitudinal direction to selectively transport atoms in either one of two internal states, $\ket{\uparrow}$ and $\ket{\downarrow}$.
In essence, two copropagating laser beams of opposite circular polarization interfere with a third, counterpropagating, linearly polarized beam.
Their interference gives rise to two standing waves of left- and right-handed polarization, whose positions are actively controlled by means of two independent optical phase-locked loops.
We obtain a residual jitter of their relative position on the order of \SI{1}{\angstrom}, which is much smaller than the longitudinal extent of the atom's wave function of $\approx\SI{20}{\nano\meter}$.
At the so-called magic wavelength $\lambda_L=\SI{866}{\nano\meter}$ of cesium atoms, the internal state $\ket{\uparrow} = \ket{F=4,m_F=4}$ interacts exclusively with the $\sigma^+$-polarized component, while $\ket{\downarrow} = \ket{F=3,m_F=3}$ predominantly interacts with the $\sigma^-$-polarized component \cite{Jaksch:1998}.
Moreover, we choose a relatively deep lattice with a depth of $U/k_B\approx\SI{80}{\micro\kelvin}$ to prevent tunneling between different sites.
Hence, atoms in the two internal states are bound to two spatially superimposed, but orthogonally polarized lattices, which can be individually shifted much like two independent optical conveyor belts:
atoms in the $\ket{\uparrow}$ and $\ket{\downarrow}$ states follow, nearly rigidly, the $\sigma^+$-polarized and $\sigma^-$-polarized standing waves, respectively.

State-dependent optical lattices have been pioneered first in the MPQ laboratories \cite{Mandel:03,Mandel:04}, demonstrating another example of Theodor W.\ Hänsch's legacy as an inspiration for future generations of experiments.
Compared to former realizations of state-dependent optical lattices, PS optical lattices have replaced the polarization control formerly based on an electrooptic modulator by a direct synthesis of light polarization, which enable arbitrary, state-dependent displacements of atoms.
Polarization synthesis is realized through rf-control of the optical phases ($\SI{0.1}{\degree}$ RMS phase jitter) of two overlapped beams with opposite circular polarization \cite{Robens:2016}.

\subsubsection{Microwave control}
We employ microwave radiation at the cesium clock frequency of $\SI{9.2}{\giga\hertz}$ to induce coherent oscillations between the two atomic hyperfine states, $\ket{\uparrow}$ and $\ket{\downarrow}$ with a Rabi frequency of $\SI{55}{\kilo\hertz}$.
Therefore, the application of the microwave radiation field for $\SI{4.5}{\micro\second}$ realizes a so-called $\pi/2$ pulse, which transforms a pure internal state into an equal superposition of $\ket{\uparrow}$ and $\ket{\downarrow}$.
In our realization of the Elitzur-Vaidman experiment, microwave $\pi/2$ pulses represent the atomic analogue of the beam splitters for photons, which are illustrated in Fig.~\ref{fig:MZI}.

\subsubsection{Non-destructive spin-state measurement}
Exploiting PS optical lattices, we devised a novel measurement method to detect the internal state of the atom in the most gentle way possible.
In quantum mechanics, least perturbative measurements are called non-destructive (or, equivalently, quantum non-demolition measurements):
the state of the object is preserved after the measurement in the eigenstate of the measured quantity corresponding to the observed outcome \cite{Haroche:2006}; a repeated measurement of the same quantity would therefore leave the state unchanged.
Conversely, the widely employed push-out method, which expels atoms in one particular internal state from the trap by applying state-selective radiation pressure \cite{Kuhr:2003}, represents a destructive measurement.

Our method is closely related to an optical Stern-Gerlach experiment \cite{Sleator:1992,Park:2002}, where spin-position entanglement is created by state-dependent light fields, and is reminiscent of the non-destructive Stern-Gerlach experiment by Dehmelt \cite{Dehmelt:1983,Dehmelt:1986}.
We realize non-destructive measurements by displacing atoms by a discrete number of lattice sites conditioned on the internal state, thereby transferring spin states to well-separated positions.
The position can be detected efficiently at a later time by fluorescence imaging under molasses illumination without any atom loss \cite{Alberti:2016};
we identify the correct position, and therefore the spin state, with $>\SI{99}{\percent}$ reliability.
The translational invariance of the optical lattice ensures that this measurement protocol constitutes a non-destructive measurement of the internal state.
Note also that it is not necessary that the position read-out immediately follows the state-dependent displacement.
The possibility to postpone the ``destructive'' fluorescence image of the atoms till the end of the evolution allows us to leave the evolution of the system minimally perturbed as required by a non-destructive spin measurement.
We use this technique for the ideal negative measurement $Q(t_2)$, in which case only one spin component at a time is displaced, leaving the other unperturbed (see Sec.~\ref{sec:measprot}).
It should also be noted that non-destructive measurements are not strictly needed in order to perform the ideal negative measurements that are instead required for the Elitzur-Vaidman experiment and the Leggett-Garg test \cite{Kwiat:1994}.
This fact is illustrated in Fig.~\ref{fig:MZI} where the photon removal represents indeed a destructive measurement, since the particle is destroyed if intercepted by the ``bomb''.

\subsection{Measurement protocols}
\label{sec:measprot}

We start each experimental sequence with, on average, $\num{1.2}$ atoms loaded into the lattice.
The loading procedure is stochastic and only atoms sitting at sufficiently separated lattice sites are considered.
Atoms are cooled to the longitudinal ground state using first molasses cooling and then microwave sideband cooling \cite{Belmechri:2013}.
With the quantization axis chosen along the lattice direction, optical pumping by a $\sigma^+$-polarized laser beam initializes $>\SI{99}{\percent}$ of the atoms in the state $\ket{\uparrow}$.

We outline in Fig.~\ref{fig:MZIa} the protocols employed to measure the two terms forming in the Leggett-Garg inequality in Eq.~\eqref{eq:LGsimp}.
On the left-hand side, we present the procedure with no ``bomb'' present, which comprises $Q(t_1)$ (preparation of the initial spin state preparation) and $Q(t_3)$ (detection of the final spin state) measurements, but not $Q(t_2)$ (the ``live bomb'' in one of the interferometer's branch):
the spin preparation is followed by a $\pi/2$ pulse, a variable waiting time, a second $\pi/2$ pulse with adjustable microwave phase $\phi$, and a non-destructive spin measurement mapping the spin state onto different positions as described in Sec.~\ref{sec:ExpApparatus}.
This sequence describes a Ramsey experiment interrogating the spin coherence of a trapped atom.
Note that this situation is fully equivalent to the unobstructed MZ interferometer of Fig.~\figref{fig:MZI}{a}, where the atomic internal states here take the place of the distinct trajectories of the photon.
Here we adjust the microwave phase $\phi$ beforehand in order to ensure the highest probability to detect $\ket{\downarrow}$ at time $t_3$, much like in the MZ interferometer one must balance the two branches to route all photons to detector D2.

Let us turn to the application of the atomic ``bomb test'' illustrated on the right-hand side of Fig.~\ref{fig:MZIa}.
After the initial microwave $\pi/2$ pulse, which puts the atom in a coherent superposition state (for a macro-realist, a stochastic mixture of both states), we spatially remove atoms, in different experiments, at one time in state $\ket{\uparrow}$ and at another time in state $\ket{\downarrow}$ by transporting them apart by seven lattice sites in about $\SI{200}{\micro\second}$.
The number of sites is chosen sufficiently large to avoid any error in the final position measurement, and therefore in the spin reconstruction.
For the Leggett-Garg test, we post-select the events where an atom is indeed found at position D1 or D2, meaning that it has not been removed at time $t_2$ and its spin has thus been measured by an ideal negative measurement $Q(t_2)$, as was argued in Sec.~\ref{sec:LeggettGarg}.
The excluded events instead correspond to the atoms removed from the Ramsey interferometer, and are tantamount to having triggered the ``bomb'' in one of the two branches of the MZ interferometer of Fig.~\figref{fig:MZI}{b}.

\section{Experimental results}

\begin{figure}[t]
	\centering
	\includegraphics[width=\columnwidth]{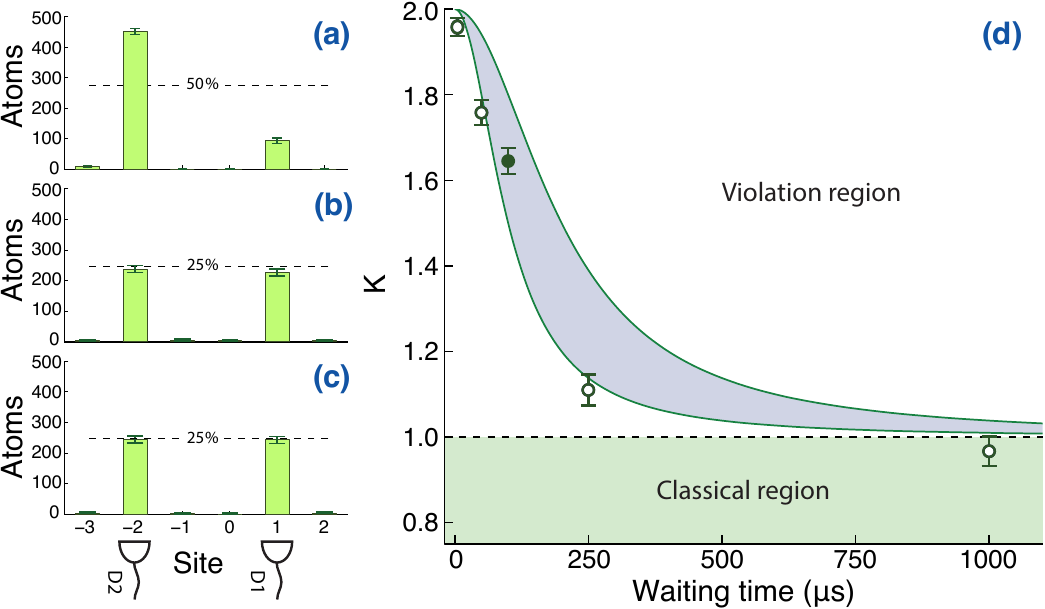}
	\caption{Experimental violation of the Leggett-Garg inequality in the quantum-to-classical transition.
    From (a) to (c), distributions at time $t_3$ of the detected atom at sites D1 and D2
	for a waiting time of $\SI{100}{\micro\second}$, corresponding to the solid point in (d) for three different protocols.
	(a) Without the $Q(t_2)$ measurement (left-hand-side protocol in Fig.~\ref{fig:MZIa}).
    (b) With the $Q(t_2)$ measurement shifting atoms in $\ket{\uparrow}$ away at time $t_2$ (right-hand-side protocol in Fig.~\ref{fig:MZIa}).
    (c) The same but with atoms in $\ket{\downarrow}$ shifted away.
    (d) Values of the Leggett-Garg correlation function $K$ of Eq. \eqref{eq:LGsimp} for increasing waiting times between the two $\pi/2$ pulses.
	Decoherence gradually suppresses the quantum behavior of the atom.
    The shaded band represents the theoretical quantum-mechanical prediction for coherence times between $\SI{75}{\micro\second}$ and $\SI{200}{\micro\second}$ caused by differential scalar light shift \cite{Kuhr:2005}.
	Percentage values are referred to the total number of interrogated atoms in each dataset.
	The vertical error bars represent $\num{1}\,\sigma$ statistical uncertainty.
	}
	\label{fig:Results}
\end{figure}

In panels Fig.~\figref{fig:Results}{a-c}, we show the raw data corresponding to our atomic realization of the Elitzur-Vaidman experiment for three different situations: (a) without the ``bomb'', (b) with the ``bomb'' removing atoms in $\ket{\uparrow}$, and (c) again with the ``bomb'' but removing atoms in $\ket{\downarrow}$.
From the dataset (a) we reconstruct the correlation function $\langle Q(t_3)\rangle_{\text{without~}Q2}$, while by merging the datasets (b) and (c) we obtain the correlation function $\langle Q(t_3)\rangle_{\text{with~}Q2}$.
The combination of these two correlation functions produce a violation of the Leggett-Garg inequality as defined in Eq.~\eqref{eq:LGsimp}.
In Fig.~\figref{fig:Results}{d} we present the recorded Leggett-Garg correlation function $K$ for different durations of the waiting time, which separates the two microwave $\pi/2$ pulses of the Ramsey interferometer.
For a minimum waiting time of $\SI{5}{\micro\second}$, we record a value of $K=1.958\pm0.033$, which violates the Leggett-Garg inequality by $\num{21}\,\sigma$.
While this value of $K$ lies very close to the decoherence-free prediction ($K=\num{2}$), the recorded values of $K$ decrease visibly for longer waiting times (i.e., increasing decoherence) till they reach the value of 1 for fully decohered spin dynamics, in fulfillment of the Leggett-Garg inequality.

In the present experiment, we attribute the main source of decoherence to scalar differential light shift \cite{Kuhr:2005}, causing inhomogeneous spin dephasing of the atoms, which in our case are thermally distributed over more than 100 vibrational levels in the directions transverse to the lattice.
Besides producing a rigorous violation of the Leggett-Garg inequality for a pseudo-spin-1/2 particle, our results show that the correlation function $K$ can be interpreted, from the point of view quantum theory, as a quantum witness of superposition states, and employed to study decoherence \textemdash one of the most basic mechanisms affecting atoms trapped in optical potentials \cite{Alberti:2014}.

\section{Conclusions}
In this paper, we have shown that the ``bomb-testing'' experiment not only provides dramatic evidence of the ``weirdness'' of quantum mechanics, as originally intended by Elitzur und Vaidman, but also can be recast in a rigorous test of the macro-realistic worldview based on the violation of the Leggett-Garg inequality. 
With our realization of the Elitzur-Vaidman experiment, we can refute the macro-realistic assumptions for individual cesium atoms with a $21\, \sigma$ statistical confidence.
While from the point of view of ``macroscopicity'', the present Leggett-Garg experiment does not improve on the results obtained with quantum walks by Robens et al.~\cite{Robens:2015}, our atomic implementation of the Elitzur-Vaidman experiment can be directly extended in the future to superposition states involving splitting distances on the macroscopic scale of a millimeter and beyond \cite{Kovachy:2015,StamperKurn:2016}.
Moreover, the analysis of the experimental results allows us to understand the exact relation between the Leggett-Garg correlation function $K$ and the interference contrast $C$ of the corresponding Ramsey interferometer, thus providing intuition about the quantum-to-classical transition of the Leggett-Garg test.

It is also worth noting that our experiment demonstrates a non-destructive measurement technique of the spin state of the atoms, where spin-position entanglement is used to transfer information from spin to position space.
This mapping technique allows us to directly read out both spin states in a Ramsey interferometer, thus avoiding the shortcomings of the widely used push-out technique, where atoms are lost after the measurement and, most importantly, the measurement outcomes must be corrected for atom losses.
Further, with the non-destructive measurement technique demonstrated here we could recycle atoms multiple times.
We finally anticipate that non-destructive spin measurements, preserving spatial coherence of atoms delocalized over several lattice sites, can find application in the realization of dissipative quantum-walk protocols \cite{Asboth:2016}.

\begin{acknowledgments}
We are very grateful to Jean-Michel Raimond for insightful discussions. We acknowledge financial support from NRW-Nachwuchsforschergruppe ``Quantenkontrolle auf der Nanoskala,'' ERC grant DQSIM, and EU Integrated Project SIQS. In addition, C.R. acknowledges support from the BCGS program and the Studienstiftung des deutschen Volkes.
\end{acknowledgments}

\begin{appendix}
\label{sec:appendix}

\section{Statistical errors}
In this work, the confidence intervals of the correlation measurements represent $1\,\sigma$ statistical uncertainty, which has been computed by fitting a Gaussian profile to the bootstrapped distribution (i.e., the distribution obtained by resampling with replacement). Independently from bootstrapping, we also computed the statistical uncertainties using Monte Carlo resampling, where the statistical errors of position distributions are estimated with binomial statistics (Clopper-Pearson method). The two estimation methods lead to consistent results. While Monte Carlo analysis requires invariant statistical properties to be valid, bootstrapping analysis remains valid also in the presence of slow drifts of experimental parameters.
The close agreement between the two statistical analyses indicates that each correlation measurement of $K$ (lasting about $\SI{120}{\minute}$) is performed under constant experimental conditions.

\section{Systematic errors}
Provided that the experiment is performed under constant experimental conditions, systematic errors do not invalidate the result of a Leggett-Garg test. In fact, if we consider the three main mechanisms that bring about systematic fluctuations:
(1) Imperfect initialization prepares $<\SI{1}{\percent}$ of the atoms in the wrong internal state. However, to derive the Leggett-Garg inequality, a statistical mixture defining the initial state is perfectly admissible.
(2) Imperfect reconstruction of the atom's position can be accounted for in terms of a noisy measurement apparatus.
(3) Spontaneous spin flips due to the finite $T_1$ time can be accounted for in terms of an additional stochastic process, which also contributes to determine the system's evolution.
We estimate that each of these three mechanisms actually affects the position distribution by $<\SI{1}{\percent}$, that is less than the statistical uncertainty.

\section{Dichotomic choice}
\label{app:dichotomic}
We verified that our system produces a violation also with a dichotomic definition of $Q(t_2)$. We performed a Ramsey sequence using two subsequent $\pi/3$ pulses. In this case, we set $Q(t_1)=1$ by preparation and designated both $Q(t_2)$ and $Q(t_3)$ as $+1$ for $\ket{\uparrow}$ and as $-1$ for $\ket{\downarrow}$.
The measured value $K=\num{1.503\pm0.051}$ is consistent with the quantum mechanical expectation of $K=\num{3/2}$.

\section{Relation with Ramsey contrast}
\label{app:RamseyContrast}
A Ramsey interference fringe is represented by the probability $p_\uparrow$ of measuring $\ket{\uparrow}$ as a function of the Ramsey phase. Assuming the fringe is centered around the average value $\overline{p}_\uparrow = 1/2$ (this is true in case of, e.g., pure spin dephasing), the contrast can be expressed as
\begin{equation}
	C = p_{\uparrow, \text{max}} - p_{\uparrow, \text{min}} = 1- 2 \hspace{1pt}p_{\uparrow, \text{min}},
\end{equation}
where $p_{\uparrow,\text{max/min}}$ is the maximum/minimum value of the fringe.
Because the two Ramsey $\pi/2$ pulses in the Elitzur-Vaidman experiment are set to have the same phase (Sec.~\ref{sec:EV_exp}), the value of $p_\uparrow$ measured when evaluating $\langle Q(t_3)Q(t_1)\rangle$ corresponds to $p_{\uparrow, \text{min}}$.
Hence, we obtain that the correlation function reads
\begin{equation}
	\langle Q(t_3)Q(t_1)\rangle = p_\uparrow - p_\downarrow = p_\uparrow - (1-p_\uparrow) = -C,
\end{equation}
which together with the definition of the LG inequality in Eq.~(\ref{eq:LG}) proves Eq.~(\ref{eq:LGRamsey}).

\end{appendix}

\bibliographystyle{apsrev4-1}
\bibliography{atomic-bomb-tester}

\begin{thebibliography}{45}%
\makeatletter
\providecommand \@ifxundefined [1]{%
 \@ifx{#1\undefined}
}%
\providecommand \@ifnum [1]{%
 \ifnum #1\expandafter \@firstoftwo
 \else \expandafter \@secondoftwo
 \fi
}%
\providecommand \@ifx [1]{%
 \ifx #1\expandafter \@firstoftwo
 \else \expandafter \@secondoftwo
 \fi
}%
\providecommand \natexlab [1]{#1}%
\providecommand \enquote  [1]{``#1''}%
\providecommand \bibnamefont  [1]{#1}%
\providecommand \bibfnamefont [1]{#1}%
\providecommand \citenamefont [1]{#1}%
\providecommand \href@noop [0]{\@secondoftwo}%
\providecommand \href [0]{\begingroup \@sanitize@url \@href}%
\providecommand \@href[1]{\@@startlink{#1}\@@href}%
\providecommand \@@href[1]{\endgroup#1\@@endlink}%
\providecommand \@sanitize@url [0]{\catcode `\\12\catcode `\$12\catcode
  `\&12\catcode `\#12\catcode `\^12\catcode `\_12\catcode `\%12\relax}%
\providecommand \@@startlink[1]{}%
\providecommand \@@endlink[0]{}%
\providecommand \url  [0]{\begingroup\@sanitize@url \@url }%
\providecommand \@url [1]{\endgroup\@href {#1}{\urlprefix }}%
\providecommand \urlprefix  [0]{URL }%
\providecommand \Eprint [0]{\href }%
\providecommand \doibase [0]{http://dx.doi.org/}%
\providecommand \selectlanguage [0]{\@gobble}%
\providecommand \bibinfo  [0]{\@secondoftwo}%
\providecommand \bibfield  [0]{\@secondoftwo}%
\providecommand \translation [1]{[#1]}%
\providecommand \BibitemOpen [0]{}%
\providecommand \bibitemStop [0]{}%
\providecommand \bibitemNoStop [0]{.\EOS\space}%
\providecommand \EOS [0]{\spacefactor3000\relax}%
\providecommand \BibitemShut  [1]{\csname bibitem#1\endcsname}%
\let\auto@bib@innerbib\@empty
%</preamble>
\bibitem [{\citenamefont {Wheeler}\ and\ \citenamefont
  {Zurek}(1983)}]{Wheeler:1983}%
  \BibitemOpen
  \bibinfo {editor} {\bibfnamefont {J.~A.}\ \bibnamefont {Wheeler}}\ and\
  \bibinfo {editor} {\bibfnamefont {W.~H.}\ \bibnamefont {Zurek}},\ eds.,\
  \href@noop {} {\emph {\bibinfo {title} {Quantum Theory and Measurement}}},\
  Princeton Series in Physics\ (\bibinfo  {publisher} {Princeton University
  Press},\ \bibinfo {address} {United States},\ \bibinfo {year}
  {1983})\BibitemShut {NoStop}%
\bibitem [{\citenamefont {Leggett}(2005)}]{Leggett:2005}%
  \BibitemOpen
  \bibfield  {author} {\bibinfo {author} {\bibfnamefont {A.~J.}\ \bibnamefont
  {Leggett}},\ }\bibfield  {title} {\enquote {\bibinfo {title} {{The quantum
  measurement problem}},}\ }\href {http://dx.doi.org/10.1126/science.1109541}
  {\bibfield  {journal} {\bibinfo  {journal} {Science}\ }\textbf {\bibinfo
  {volume} {307}},\ \bibinfo {pages} {871} (\bibinfo {year}
  {2005})}\BibitemShut {NoStop}%
\bibitem [{\citenamefont {L{\"u}ders}(1951)}]{Lueders:1951}%
  \BibitemOpen
  \bibfield  {author} {\bibinfo {author} {\bibfnamefont {G.}~\bibnamefont
  {L{\"u}ders}},\ }\bibfield  {title} {\enquote {\bibinfo {title} {{Concerning
  the state-change due to the measurement process}},}\ }\href
  {http://dx.doi.org/10.1002/andp.200610207} {\bibfield  {journal} {\bibinfo
  {journal} {Ann. Phys. (Leipzig)}\ }\textbf {\bibinfo {volume} {8}},\ \bibinfo
  {pages} {332} (\bibinfo {year} {1951})}\BibitemShut {NoStop}%
\bibitem [{\citenamefont {Bassi}\ \emph {et~al.}(2013)\citenamefont {Bassi},
  \citenamefont {Lochan}, \citenamefont {Satin}, \citenamefont {Singh},\ and\
  \citenamefont {Ulbricht}}]{Bassi:2013}%
  \BibitemOpen
  \bibfield  {author} {\bibinfo {author} {\bibfnamefont {A.}~\bibnamefont
  {Bassi}}, \bibinfo {author} {\bibfnamefont {K.}~\bibnamefont {Lochan}},
  \bibinfo {author} {\bibfnamefont {S.}~\bibnamefont {Satin}}, \bibinfo
  {author} {\bibfnamefont {T.~P.}\ \bibnamefont {Singh}}, \ and\ \bibinfo
  {author} {\bibfnamefont {H.}~\bibnamefont {Ulbricht}},\ }\bibfield  {title}
  {\enquote {\bibinfo {title} {{Models of wave-function collapse, underlying
  theories, and experimental tests}},}\ }\href
  {http://dx.doi.org/10.1103/RevModPhys.85.471} {\bibfield  {journal} {\bibinfo
   {journal} {Rev. Mod. Phys.}\ }\textbf {\bibinfo {volume} {85}},\ \bibinfo
  {pages} {471} (\bibinfo {year} {2013})}\BibitemShut {NoStop}%
\bibitem [{\citenamefont {Ghirardi}\ \emph {et~al.}(1986)\citenamefont
  {Ghirardi}, \citenamefont {Rimini},\ and\ \citenamefont
  {Weber}}]{Ghirardi:1986}%
  \BibitemOpen
  \bibfield  {author} {\bibinfo {author} {\bibfnamefont {G.}~\bibnamefont
  {Ghirardi}}, \bibinfo {author} {\bibfnamefont {A.}~\bibnamefont {Rimini}}, \
  and\ \bibinfo {author} {\bibfnamefont {T.}~\bibnamefont {Weber}},\ }\bibfield
   {title} {\enquote {\bibinfo {title} {{Unified dynamics for microscopic and
  macroscopic systems.}}}\ }\href {http://dx.doi.org/10.1103/PhysRevD.34.470}
  {\bibfield  {journal} {\bibinfo  {journal} {Phys. Rev. D}\ }\textbf {\bibinfo
  {volume} {34}},\ \bibinfo {pages} {470} (\bibinfo {year} {1986})}\BibitemShut
  {NoStop}%
\bibitem [{\citenamefont {Di{\'o}si}(1989)}]{Diosi:1989}%
  \BibitemOpen
  \bibfield  {author} {\bibinfo {author} {\bibfnamefont {L.}~\bibnamefont
  {Di{\'o}si}},\ }\bibfield  {title} {\enquote {\bibinfo {title} {{Models for
  universal reduction of macroscopic quantum fluctuations}},}\ }\href
  {http://dx.doi.org/10.1103/PhysRevA.40.1165} {\bibfield  {journal} {\bibinfo
  {journal} {Phys. Rev. A}\ }\textbf {\bibinfo {volume} {40}},\ \bibinfo
  {pages} {1165} (\bibinfo {year} {1989})}\BibitemShut {NoStop}%
\bibitem [{\citenamefont {Pikovski}\ \emph {et~al.}(2015)\citenamefont
  {Pikovski}, \citenamefont {Zych}, \citenamefont {Costa},\ and\ \citenamefont
  {Brukner}}]{Pikovski:2015}%
  \BibitemOpen
  \bibfield  {author} {\bibinfo {author} {\bibfnamefont {I.}~\bibnamefont
  {Pikovski}}, \bibinfo {author} {\bibfnamefont {M.}~\bibnamefont {Zych}},
  \bibinfo {author} {\bibfnamefont {F.}~\bibnamefont {Costa}}, \ and\ \bibinfo
  {author} {\bibfnamefont {C.}~\bibnamefont {Brukner}},\ }\bibfield  {title}
  {\enquote {\bibinfo {title} {{Universal decoherence due to gravitational time
  dilation}},}\ }\href {http://dx.doi.org/10.1038/nphys3366} {\bibfield
  {journal} {\bibinfo  {journal} {Nature Phys.}\ }\textbf {\bibinfo {volume}
  {11}},\ \bibinfo {pages} {668} (\bibinfo {year} {2015})}\BibitemShut
  {NoStop}%
\bibitem [{\citenamefont {Emary}\ \emph {et~al.}(2013)\citenamefont {Emary},
  \citenamefont {Lambert},\ and\ \citenamefont {Nori}}]{Emary:2013}%
  \BibitemOpen
  \bibfield  {author} {\bibinfo {author} {\bibfnamefont {C.}~\bibnamefont
  {Emary}}, \bibinfo {author} {\bibfnamefont {N.}~\bibnamefont {Lambert}}, \
  and\ \bibinfo {author} {\bibfnamefont {F.}~\bibnamefont {Nori}},\ }\bibfield
  {title} {\enquote {\bibinfo {title} {{Leggett-Garg inequalities.}}}\ }\href
  {http://dx.doi.org/10.1088/0034-4885/77/1/016001} {\bibfield  {journal}
  {\bibinfo  {journal} {Rep. Prog. Phys.}\ }\textbf {\bibinfo {volume} {77}},\
  \bibinfo {pages} {016001} (\bibinfo {year} {2013})}\BibitemShut {NoStop}%
\bibitem [{\citenamefont {Leggett}\ and\ \citenamefont
  {Garg}(1985)}]{Leggett:1985}%
  \BibitemOpen
  \bibfield  {author} {\bibinfo {author} {\bibfnamefont {A.~J.}\ \bibnamefont
  {Leggett}}\ and\ \bibinfo {author} {\bibfnamefont {A.}~\bibnamefont {Garg}},\
  }\bibfield  {title} {\enquote {\bibinfo {title} {{Quantum mechanics versus
  macroscopic realism: Is the flux there when nobody looks?}}}\ }\href
  {http://dx.doi.org/10.1103/PhysRevLett.54.857} {\bibfield  {journal}
  {\bibinfo  {journal} {Phys. Rev. Lett.}\ }\textbf {\bibinfo {volume} {54}},\
  \bibinfo {pages} {857} (\bibinfo {year} {1985})}\BibitemShut {NoStop}%
\bibitem [{\citenamefont {Knee}\ \emph {et~al.}(2012)\citenamefont {Knee},
  \citenamefont {Simmons}, \citenamefont {Gauger}, \citenamefont {Morton},
  \citenamefont {Riemann}, \citenamefont {Abrosimov}, \citenamefont {Becker},
  \citenamefont {Pohl}, \citenamefont {Itoh}, \citenamefont {Thewalt},
  \citenamefont {Briggs},\ and\ \citenamefont {Benjamin}}]{Briggs:2012}%
  \BibitemOpen
  \bibfield  {author} {\bibinfo {author} {\bibfnamefont {G.~C.}\ \bibnamefont
  {Knee}}, \bibinfo {author} {\bibfnamefont {S.}~\bibnamefont {Simmons}},
  \bibinfo {author} {\bibfnamefont {E.~M.}\ \bibnamefont {Gauger}}, \bibinfo
  {author} {\bibfnamefont {J.~J.~L.}\ \bibnamefont {Morton}}, \bibinfo {author}
  {\bibfnamefont {H.}~\bibnamefont {Riemann}}, \bibinfo {author} {\bibfnamefont
  {N.~V.}\ \bibnamefont {Abrosimov}}, \bibinfo {author} {\bibfnamefont
  {P.}~\bibnamefont {Becker}}, \bibinfo {author} {\bibfnamefont {H.-J.}\
  \bibnamefont {Pohl}}, \bibinfo {author} {\bibfnamefont {K.~M.}\ \bibnamefont
  {Itoh}}, \bibinfo {author} {\bibfnamefont {M.~L.~W.}\ \bibnamefont
  {Thewalt}}, \bibinfo {author} {\bibfnamefont {G.~A.~D.}\ \bibnamefont
  {Briggs}}, \ and\ \bibinfo {author} {\bibfnamefont {S.~C.}\ \bibnamefont
  {Benjamin}},\ }\bibfield  {title} {\enquote {\bibinfo {title} {{Violation of
  a Leggett-Garg inequality with ideal non-invasive measurements}},}\ }\href
  {http://dx.doi.org/10.1038/ncomms1614} {\bibfield  {journal} {\bibinfo
  {journal} {Nat. Comm.}\ }\textbf {\bibinfo {volume} {3}},\ \bibinfo {pages}
  {606} (\bibinfo {year} {2012})}\BibitemShut {NoStop}%
\bibitem [{\citenamefont {Robens}\ \emph {et~al.}(2015)\citenamefont {Robens},
  \citenamefont {Alt}, \citenamefont {Meschede}, \citenamefont {Emary},\ and\
  \citenamefont {Alberti}}]{Robens:2015}%
  \BibitemOpen
  \bibfield  {author} {\bibinfo {author} {\bibfnamefont {C.}~\bibnamefont
  {Robens}}, \bibinfo {author} {\bibfnamefont {W.}~\bibnamefont {Alt}},
  \bibinfo {author} {\bibfnamefont {D.}~\bibnamefont {Meschede}}, \bibinfo
  {author} {\bibfnamefont {C.}~\bibnamefont {Emary}}, \ and\ \bibinfo {author}
  {\bibfnamefont {A.}~\bibnamefont {Alberti}},\ }\bibfield  {title} {\enquote
  {\bibinfo {title} {{Ideal Negative Measurements in Quantum Walks Disprove
  Theories Based on Classical Trajectories}},}\ }\href
  {http://dx.doi.org/10.1103/PhysRevX.5.011003} {\bibfield  {journal} {\bibinfo
   {journal} {Phys. Rev. X}\ }\textbf {\bibinfo {volume} {5}},\ \bibinfo
  {pages} {011003} (\bibinfo {year} {2015})}\BibitemShut {NoStop}%
\bibitem [{\citenamefont {Leggett}(2002)}]{Leggett:2002}%
  \BibitemOpen
  \bibfield  {author} {\bibinfo {author} {\bibfnamefont {A.~J.}\ \bibnamefont
  {Leggett}},\ }\bibfield  {title} {\enquote {\bibinfo {title} {{Testing the
  limits of quantum mechanics: motivation, state of play, prospects}},}\ }\href
  {http://dx.doi.org/10.1088/0953-8984/14/15/201} {\bibfield  {journal}
  {\bibinfo  {journal} {J. Phys.: Condens. Matter}\ }\textbf {\bibinfo {volume}
  {14}},\ \bibinfo {pages} {415} (\bibinfo {year} {2002})}\BibitemShut
  {NoStop}%
\bibitem [{\citenamefont {Arndt}\ and\ \citenamefont
  {Hornberger}(2014)}]{Arndt:2014}%
  \BibitemOpen
  \bibfield  {author} {\bibinfo {author} {\bibfnamefont {M.}~\bibnamefont
  {Arndt}}\ and\ \bibinfo {author} {\bibfnamefont {K.}~\bibnamefont
  {Hornberger}},\ }\bibfield  {title} {\enquote {\bibinfo {title} {{Testing the
  limits of quantum mechanical superpositions}},}\ }\href
  {http://dx.doi.org/10.1038/nphys2863} {\bibfield  {journal} {\bibinfo
  {journal} {Nature Phys.}\ }\textbf {\bibinfo {volume} {10}},\ \bibinfo
  {pages} {271} (\bibinfo {year} {2014})}\BibitemShut {NoStop}%
\bibitem [{\citenamefont {Nimmrichter}\ and\ \citenamefont
  {Hornberger}(2013)}]{Nimmrichter:2013}%
  \BibitemOpen
  \bibfield  {author} {\bibinfo {author} {\bibfnamefont {S.}~\bibnamefont
  {Nimmrichter}}\ and\ \bibinfo {author} {\bibfnamefont {K.}~\bibnamefont
  {Hornberger}},\ }\bibfield  {title} {\enquote {\bibinfo {title}
  {{Macroscopicity of Mechanical Quantum Superposition States}},}\ }\href
  {http://dx.doi.org/10.1103/PhysRevLett.110.160403} {\bibfield  {journal}
  {\bibinfo  {journal} {Phys. Rev. Lett.}\ }\textbf {\bibinfo {volume} {110}},\
  \bibinfo {pages} {160403} (\bibinfo {year} {2013})}\BibitemShut {NoStop}%
\bibitem [{\citenamefont {Ghirardi}(2007)}]{Ghirardi:2007}%
  \BibitemOpen
  \bibfield  {author} {\bibinfo {author} {\bibfnamefont {G.}~\bibnamefont
  {Ghirardi}},\ }\href@noop {} {\emph {\bibinfo {title} {Sneaking a Look at
  God's Cards, Revised Edition: Unraveling the Mysteries of Quantum
  Mechanics}}}\ (\bibinfo  {publisher} {Princeton University Press},\ \bibinfo
  {address} {United States},\ \bibinfo {year} {2007})\BibitemShut {NoStop}%
\bibitem [{\citenamefont {Elitzur}\ and\ \citenamefont
  {Vaidman}(1993)}]{EV:1993}%
  \BibitemOpen
  \bibfield  {author} {\bibinfo {author} {\bibfnamefont {A.~C.}\ \bibnamefont
  {Elitzur}}\ and\ \bibinfo {author} {\bibfnamefont {L.}~\bibnamefont
  {Vaidman}},\ }\bibfield  {title} {\enquote {\bibinfo {title} {{Quantum
  mechanical interaction-free measurements}},}\ }\href
  {http://dx.doi.org/10.1007/BF00736012} {\bibfield  {journal} {\bibinfo
  {journal} {Found. Phys.}\ }\textbf {\bibinfo {volume} {23}},\ \bibinfo
  {pages} {987} (\bibinfo {year} {1993})}\BibitemShut {NoStop}%
\bibitem [{\citenamefont {Kwiat}\ \emph
  {et~al.}(1995{\natexlab{a}})\citenamefont {Kwiat}, \citenamefont
  {Weinfurter}, \citenamefont {Herzog}, \citenamefont {Zeilinger},\ and\
  \citenamefont {Kasevich}}]{Kwiat:1995}%
  \BibitemOpen
  \bibfield  {author} {\bibinfo {author} {\bibfnamefont {P.}~\bibnamefont
  {Kwiat}}, \bibinfo {author} {\bibfnamefont {H.}~\bibnamefont {Weinfurter}},
  \bibinfo {author} {\bibfnamefont {T.}~\bibnamefont {Herzog}}, \bibinfo
  {author} {\bibfnamefont {A.}~\bibnamefont {Zeilinger}}, \ and\ \bibinfo
  {author} {\bibfnamefont {M.~A.}\ \bibnamefont {Kasevich}},\ }\bibfield
  {title} {\enquote {\bibinfo {title} {{Interaction-Free Measurement}},}\
  }\href {http://dx.doi.org/10.1103/PhysRevLett.74.4763} {\bibfield  {journal}
  {\bibinfo  {journal} {Phys. Rev. Lett.}\ }\textbf {\bibinfo {volume} {74}},\
  \bibinfo {pages} {4763} (\bibinfo {year} {1995}{\natexlab{a}})}\BibitemShut
  {NoStop}%
\bibitem [{\citenamefont {Hafner}\ and\ \citenamefont
  {Summhammer}(1997)}]{Hafner:1997}%
  \BibitemOpen
  \bibfield  {author} {\bibinfo {author} {\bibfnamefont {M.}~\bibnamefont
  {Hafner}}\ and\ \bibinfo {author} {\bibfnamefont {J.}~\bibnamefont
  {Summhammer}},\ }\bibfield  {title} {\enquote {\bibinfo {title} {{Experiment
  on interaction-free measurement in neutron interferometry}},}\ }\href
  {http://dx.doi.org/10.1016/S0375-9601(97)00696-8} {\bibfield  {journal}
  {\bibinfo  {journal} {Phys. Lett. A}\ }\textbf {\bibinfo {volume} {235}},\
  \bibinfo {pages} {563} (\bibinfo {year} {1997})}\BibitemShut {NoStop}%
\bibitem [{Wil()}]{Wile}%
  \BibitemOpen
  \bibinfo {note} {For Wile's safety, the avalanche photodiode must be cooled
  to near absolute zero in order to suppress dark counts.}\BibitemShut {Stop}%
\bibitem [{\citenamefont {Vaidman}(2003)}]{Vaidman:2003}%
  \BibitemOpen
  \bibfield  {author} {\bibinfo {author} {\bibfnamefont {L.}~\bibnamefont
  {Vaidman}},\ }\bibfield  {title} {\enquote {\bibinfo {title} {{The Meaning of
  the Interaction-Free Measurements}},}\ }\href
  {http://dx.doi.org/10.1023/A:1023767716236} {\bibfield  {journal} {\bibinfo
  {journal} {Found. Phys.}\ }\textbf {\bibinfo {volume} {33}},\ \bibinfo
  {pages} {491} (\bibinfo {year} {2003})}\BibitemShut {NoStop}%
\bibitem [{\citenamefont {Kwiat}\ \emph
  {et~al.}(1995{\natexlab{b}})\citenamefont {Kwiat}, \citenamefont
  {Weinfurter}, \citenamefont {Herzog}, \citenamefont {Zeilinger},\ and\
  \citenamefont {Kasevich}}]{Kwiat:1994}%
  \BibitemOpen
  \bibfield  {author} {\bibinfo {author} {\bibfnamefont {P.}~\bibnamefont
  {Kwiat}}, \bibinfo {author} {\bibfnamefont {H.}~\bibnamefont {Weinfurter}},
  \bibinfo {author} {\bibfnamefont {T.}~\bibnamefont {Herzog}}, \bibinfo
  {author} {\bibfnamefont {A.}~\bibnamefont {Zeilinger}}, \ and\ \bibinfo
  {author} {\bibfnamefont {M.}~\bibnamefont {Kasevich}},\ }\bibfield  {title}
  {\enquote {\bibinfo {title} {{Experimental Realization of
  ``Interaction-Free'' Measurements}},}\ }in\ \href@noop {} {\emph {\bibinfo
  {booktitle} {{Fundamental Problems in Quantum Theory: A Conference Held in
  Honor ofProfessor John A. Wheeler}}}},\ Vol.\ \bibinfo {volume} {755},\
  \bibinfo {editor} {edited by\ \bibinfo {editor} {\bibfnamefont {N.~Y.~A.}\
  \bibnamefont {of~Sciences}}}\ (\bibinfo {address} {New York},\ \bibinfo
  {year} {1995})\ p.\ \bibinfo {pages} {129}\BibitemShut {NoStop}%
\bibitem [{\citenamefont {Kwiat}\ \emph {et~al.}(1999)\citenamefont {Kwiat},
  \citenamefont {White}, \citenamefont {Mitchell}, \citenamefont {Nairz},
  \citenamefont {Weihs}, \citenamefont {Weinfurter},\ and\ \citenamefont
  {Zeilinger}}]{Kwiat:1999}%
  \BibitemOpen
  \bibfield  {author} {\bibinfo {author} {\bibfnamefont {P.~G.}\ \bibnamefont
  {Kwiat}}, \bibinfo {author} {\bibfnamefont {A.~G.}\ \bibnamefont {White}},
  \bibinfo {author} {\bibfnamefont {J.~R.}\ \bibnamefont {Mitchell}}, \bibinfo
  {author} {\bibfnamefont {O.}~\bibnamefont {Nairz}}, \bibinfo {author}
  {\bibfnamefont {G.}~\bibnamefont {Weihs}}, \bibinfo {author} {\bibfnamefont
  {H.}~\bibnamefont {Weinfurter}}, \ and\ \bibinfo {author} {\bibfnamefont
  {A.}~\bibnamefont {Zeilinger}},\ }\bibfield  {title} {\enquote {\bibinfo
  {title} {{High-Efficiency Quantum Interrogation Measurements via the Quantum
  Zeno Effect}},}\ }\href {http://dx.doi.org/10.1103/PhysRevLett.83.4725}
  {\bibfield  {journal} {\bibinfo  {journal} {Phys. Rev. Lett.}\ }\textbf
  {\bibinfo {volume} {83}},\ \bibinfo {pages} {4725} (\bibinfo {year}
  {1999})}\BibitemShut {NoStop}%
\bibitem [{\citenamefont {Emary}(2013)}]{Emary2013a}%
  \BibitemOpen
  \bibfield  {author} {\bibinfo {author} {\bibfnamefont {C.}~\bibnamefont
  {Emary}},\ }\bibfield  {title} {\enquote {\bibinfo {title} {{Decoherence and
  maximal violations of the Leggett-Garg inequality}},}\ }\href
  {http://dx.doi.org/10.1103/PhysRevA.87.032106} {\bibfield  {journal}
  {\bibinfo  {journal} {Phys. Rev. A}\ }\textbf {\bibinfo {volume} {87}},\
  \bibinfo {pages} {032106} (\bibinfo {year} {2013})}\BibitemShut {NoStop}%
\bibitem [{\citenamefont {Wilde}\ and\ \citenamefont
  {Mizel}(2012)}]{Wilde:2012}%
  \BibitemOpen
  \bibfield  {author} {\bibinfo {author} {\bibfnamefont {M.~M.}\ \bibnamefont
  {Wilde}}\ and\ \bibinfo {author} {\bibfnamefont {A.}~\bibnamefont {Mizel}},\
  }\bibfield  {title} {\enquote {\bibinfo {title} {{Addressing the Clumsiness
  Loophole in a Leggett-Garg Test of Macrorealism}},}\ }\href
  {http://dx.doi.org/10.1007/s10701-011-9598-4} {\bibfield  {journal} {\bibinfo
   {journal} {Found. Phys.}\ }\textbf {\bibinfo {volume} {42}},\ \bibinfo
  {pages} {256} (\bibinfo {year} {2012})}\BibitemShut {NoStop}%
\bibitem [{\citenamefont {Budroni}\ and\ \citenamefont
  {Emary}(2013)}]{Budroni:2013}%
  \BibitemOpen
  \bibfield  {author} {\bibinfo {author} {\bibfnamefont {C.}~\bibnamefont
  {Budroni}}\ and\ \bibinfo {author} {\bibfnamefont {C.}~\bibnamefont
  {Emary}},\ }\bibfield  {title} {\enquote {\bibinfo {title} {{Temporal quantum
  correlations and Leggett-Garg inequalities in multi-level systems}},}\
  }\href@noop {} {\bibfield  {journal} {\bibinfo  {journal} {preprint}\
  }\textbf {\bibinfo {volume} {{}}} (\bibinfo {year} {2013})},\ \Eprint
  {http://arxiv.org/abs/arXiv:1309.3678} {arXiv:1309.3678} \BibitemShut
  {NoStop}%
\bibitem [{\citenamefont {Schild}\ and\ \citenamefont
  {Emary}(2015)}]{Schild:2015}%
  \BibitemOpen
  \bibfield  {author} {\bibinfo {author} {\bibfnamefont {G.}~\bibnamefont
  {Schild}}\ and\ \bibinfo {author} {\bibfnamefont {C.}~\bibnamefont {Emary}},\
  }\bibfield  {title} {\enquote {\bibinfo {title} {{Maximum violations of the
  quantum-witness equality}},}\ }\href
  {http://dx.doi.org/10.1103/PhysRevA.92.032101} {\bibfield  {journal}
  {\bibinfo  {journal} {Phys. Rev. A}\ }\textbf {\bibinfo {volume} {92}},\
  \bibinfo {pages} {032101} (\bibinfo {year} {2015})}\BibitemShut {NoStop}%
\bibitem [{\citenamefont {Li}\ \emph {et~al.}(2012)\citenamefont {Li},
  \citenamefont {Lambert}, \citenamefont {Chen}, \citenamefont {Chen},\ and\
  \citenamefont {Nori}}]{Che-Ming:2012}%
  \BibitemOpen
  \bibfield  {author} {\bibinfo {author} {\bibfnamefont {C.-M.}\ \bibnamefont
  {Li}}, \bibinfo {author} {\bibfnamefont {N.}~\bibnamefont {Lambert}},
  \bibinfo {author} {\bibfnamefont {Y.-N.}\ \bibnamefont {Chen}}, \bibinfo
  {author} {\bibfnamefont {G.-Y.}\ \bibnamefont {Chen}}, \ and\ \bibinfo
  {author} {\bibfnamefont {F.}~\bibnamefont {Nori}},\ }\bibfield  {title}
  {\enquote {\bibinfo {title} {{Witnessing Quantum Coherence: from solid-state
  to biological systems}},}\ }\href {http://dx.doi.org/10.1038/srep00885}
  {\bibfield  {journal} {\bibinfo  {journal} {Sci. Rep.}\ }\textbf {\bibinfo
  {volume} {2}},\ \bibinfo {pages} {885} (\bibinfo {year} {2012})}\BibitemShut
  {NoStop}%
\bibitem [{\citenamefont {Kofler}\ and\ \citenamefont
  {Brukner}(2013)}]{Brukner:2013}%
  \BibitemOpen
  \bibfield  {author} {\bibinfo {author} {\bibfnamefont {J.}~\bibnamefont
  {Kofler}}\ and\ \bibinfo {author} {\bibfnamefont {C.}~\bibnamefont
  {Brukner}},\ }\bibfield  {title} {\enquote {\bibinfo {title} {{Condition for
  macroscopic realism beyond the Leggett-Garg inequalities}},}\ }\href
  {http://dx.doi.org/10.1103/PhysRevA.87.052115} {\bibfield  {journal}
  {\bibinfo  {journal} {Phys. Rev. A}\ }\textbf {\bibinfo {volume} {87}},\
  \bibinfo {pages} {052115} (\bibinfo {year} {2013})}\BibitemShut {NoStop}%
\bibitem [{\citenamefont {Robens}\ \emph {et~al.}(2016)\citenamefont {Robens},
  \citenamefont {Zopes}, \citenamefont {Alt}, \citenamefont {Brakhane},
  \citenamefont {Meschede},\ and\ \citenamefont {Alberti}}]{Robens:2016}%
  \BibitemOpen
  \bibfield  {author} {\bibinfo {author} {\bibfnamefont {C.}~\bibnamefont
  {Robens}}, \bibinfo {author} {\bibfnamefont {J.}~\bibnamefont {Zopes}},
  \bibinfo {author} {\bibfnamefont {W.}~\bibnamefont {Alt}}, \bibinfo {author}
  {\bibfnamefont {S.}~\bibnamefont {Brakhane}}, \bibinfo {author}
  {\bibfnamefont {D.}~\bibnamefont {Meschede}}, \ and\ \bibinfo {author}
  {\bibfnamefont {A.}~\bibnamefont {Alberti}},\ }\bibfield  {title} {\enquote
  {\bibinfo {title} {{Low-entropy states of neutral atoms in
  polarization-synthesized optical lattices}},}\ }\href
  {https://arxiv.org/abs/1608.02410} {\bibfield  {journal} {\bibinfo  {journal}
  {arXiv:1608.02410 [quant-ph]}\ } (\bibinfo {year} {2016})}\BibitemShut
  {NoStop}%
\bibitem [{\citenamefont {Jaksch}\ \emph {et~al.}(1998)\citenamefont {Jaksch},
  \citenamefont {Bruder}, \citenamefont {Cirac}, \citenamefont {Gardiner},\
  and\ \citenamefont {Zoller}}]{Jaksch:1998}%
  \BibitemOpen
  \bibfield  {author} {\bibinfo {author} {\bibfnamefont {D.}~\bibnamefont
  {Jaksch}}, \bibinfo {author} {\bibfnamefont {C.}~\bibnamefont {Bruder}},
  \bibinfo {author} {\bibfnamefont {J.~I.}\ \bibnamefont {Cirac}}, \bibinfo
  {author} {\bibfnamefont {C.~W.}\ \bibnamefont {Gardiner}}, \ and\ \bibinfo
  {author} {\bibfnamefont {P.}~\bibnamefont {Zoller}},\ }\bibfield  {title}
  {\enquote {\bibinfo {title} {{Cold Bosonic Atoms in Optical Lattices}},}\
  }\href {http://dx.doi.org/10.1103/PhysRevLett.81.3108} {\bibfield  {journal}
  {\bibinfo  {journal} {Phys. Rev. Lett.}\ }\textbf {\bibinfo {volume} {81}},\
  \bibinfo {pages} {3108} (\bibinfo {year} {1998})}\BibitemShut {NoStop}%
\bibitem [{\citenamefont {Mandel}\ \emph {et~al.}(2003)\citenamefont {Mandel},
  \citenamefont {Greiner}, \citenamefont {Widera}, \citenamefont {Rom},
  \citenamefont {H{\"a}nsch},\ and\ \citenamefont {Bloch}}]{Mandel:03}%
  \BibitemOpen
  \bibfield  {author} {\bibinfo {author} {\bibfnamefont {O.}~\bibnamefont
  {Mandel}}, \bibinfo {author} {\bibfnamefont {M.}~\bibnamefont {Greiner}},
  \bibinfo {author} {\bibfnamefont {A.}~\bibnamefont {Widera}}, \bibinfo
  {author} {\bibfnamefont {T.}~\bibnamefont {Rom}}, \bibinfo {author}
  {\bibfnamefont {T.~W.}\ \bibnamefont {H{\"a}nsch}}, \ and\ \bibinfo {author}
  {\bibfnamefont {I.}~\bibnamefont {Bloch}},\ }\bibfield  {title} {\enquote
  {\bibinfo {title} {Coherent transport of neutral atoms in spin-dependent
  optical lattice potentials},}\ }\href
  {http://dx.doi.org/10.1103/PhysRevLett.91.010407} {\bibfield  {journal}
  {\bibinfo  {journal} {Phys. Rev. Lett.}\ }\textbf {\bibinfo {volume} {91}},\
  \bibinfo {pages} {010407} (\bibinfo {year} {2003})}\BibitemShut {NoStop}%
\bibitem [{\citenamefont {Mandel}\ \emph {et~al.}(2004)\citenamefont {Mandel},
  \citenamefont {Greiner}, \citenamefont {Widera}, \citenamefont {Rom},
  \citenamefont {H\"ansch},\ and\ \citenamefont {Bloch}}]{Mandel:04}%
  \BibitemOpen
  \bibfield  {author} {\bibinfo {author} {\bibfnamefont {O.}~\bibnamefont
  {Mandel}}, \bibinfo {author} {\bibfnamefont {M.}~\bibnamefont {Greiner}},
  \bibinfo {author} {\bibfnamefont {A.}~\bibnamefont {Widera}}, \bibinfo
  {author} {\bibfnamefont {T.}~\bibnamefont {Rom}}, \bibinfo {author}
  {\bibfnamefont {T.}~\bibnamefont {H\"ansch}}, \ and\ \bibinfo {author}
  {\bibfnamefont {I.}~\bibnamefont {Bloch}},\ }\bibfield  {title} {\enquote
  {\bibinfo {title} {Controlled collisions for multi-particle entanglement of
  optically trapped atoms},}\ }\href {http://dx.doi.org/10.1038/nature02008}
  {\bibfield  {journal} {\bibinfo  {journal} {Nature}\ }\textbf {\bibinfo
  {volume} {425}},\ \bibinfo {pages} {937} (\bibinfo {year}
  {2004})}\BibitemShut {NoStop}%
\bibitem [{\citenamefont {Haroche}\ and\ \citenamefont
  {Raimond}(2006)}]{Haroche:2006}%
  \BibitemOpen
  \bibfield  {author} {\bibinfo {author} {\bibfnamefont {S.}~\bibnamefont
  {Haroche}}\ and\ \bibinfo {author} {\bibfnamefont {J.~M.}\ \bibnamefont
  {Raimond}},\ }\href@noop {} {\emph {\bibinfo {title} {{Exploring the Quantum:
  Atoms, Cavities, and Photons}}}}\ (\bibinfo  {publisher} {Oxford University
  Press},\ \bibinfo {address} {New York},\ \bibinfo {year} {2006})\BibitemShut
  {NoStop}%
\bibitem [{\citenamefont {Kuhr}\ \emph {et~al.}(2003)\citenamefont {Kuhr},
  \citenamefont {Alt}, \citenamefont {Schrader}, \citenamefont {Dotsenko},
  \citenamefont {Miroshnychenko}, \citenamefont {Rosenfeld}, \citenamefont
  {Khudaverdyan}, \citenamefont {Gomer}, \citenamefont {Rauschenbeutel},\ and\
  \citenamefont {Meschede}}]{Kuhr:2003}%
  \BibitemOpen
  \bibfield  {author} {\bibinfo {author} {\bibfnamefont {S.}~\bibnamefont
  {Kuhr}}, \bibinfo {author} {\bibfnamefont {W.}~\bibnamefont {Alt}}, \bibinfo
  {author} {\bibfnamefont {D.}~\bibnamefont {Schrader}}, \bibinfo {author}
  {\bibfnamefont {I.}~\bibnamefont {Dotsenko}}, \bibinfo {author}
  {\bibfnamefont {Y.}~\bibnamefont {Miroshnychenko}}, \bibinfo {author}
  {\bibfnamefont {W.}~\bibnamefont {Rosenfeld}}, \bibinfo {author}
  {\bibfnamefont {M.}~\bibnamefont {Khudaverdyan}}, \bibinfo {author}
  {\bibfnamefont {V.}~\bibnamefont {Gomer}}, \bibinfo {author} {\bibfnamefont
  {A.}~\bibnamefont {Rauschenbeutel}}, \ and\ \bibinfo {author} {\bibfnamefont
  {D.}~\bibnamefont {Meschede}},\ }\bibfield  {title} {\enquote {\bibinfo
  {title} {{Coherence Properties and Quantum State Transportation in an Optical
  Conveyor Belt}},}\ }\href {http://dx.doi.org/10.1103/PhysRevLett.91.213002}
  {\bibfield  {journal} {\bibinfo  {journal} {Phys. Rev. Lett.}\ }\textbf
  {\bibinfo {volume} {91}},\ \bibinfo {pages} {213002} (\bibinfo {year}
  {2003})}\BibitemShut {NoStop}%
\bibitem [{\citenamefont {Sleator}\ \emph {et~al.}(1992)\citenamefont
  {Sleator}, \citenamefont {Pfau}, \citenamefont {Balykin}, \citenamefont
  {Carnal},\ and\ \citenamefont {Mlynek}}]{Sleator:1992}%
  \BibitemOpen
  \bibfield  {author} {\bibinfo {author} {\bibfnamefont {T.}~\bibnamefont
  {Sleator}}, \bibinfo {author} {\bibfnamefont {T.}~\bibnamefont {Pfau}},
  \bibinfo {author} {\bibfnamefont {V.}~\bibnamefont {Balykin}}, \bibinfo
  {author} {\bibfnamefont {O.}~\bibnamefont {Carnal}}, \ and\ \bibinfo {author}
  {\bibfnamefont {J.}~\bibnamefont {Mlynek}},\ }\bibfield  {title} {\enquote
  {\bibinfo {title} {{Experimental demonstration of the optical Stern-Gerlach
  effect}},}\ }\href {http://dx.doi.org/10.1103/physrevlett.68.1996} {\bibfield
   {journal} {\bibinfo  {journal} {Phys. Rev. Lett.}\ }\textbf {\bibinfo
  {volume} {68}},\ \bibinfo {pages} {1996} (\bibinfo {year}
  {1992})}\BibitemShut {NoStop}%
\bibitem [{\citenamefont {Park}\ \emph {et~al.}(2002)\citenamefont {Park},
  \citenamefont {Kim}, \citenamefont {Song},\ and\ \citenamefont
  {Cho}}]{Park:2002}%
  \BibitemOpen
  \bibfield  {author} {\bibinfo {author} {\bibfnamefont {C.~Y.}\ \bibnamefont
  {Park}}, \bibinfo {author} {\bibfnamefont {J.~Y.}\ \bibnamefont {Kim}},
  \bibinfo {author} {\bibfnamefont {J.~M.}\ \bibnamefont {Song}}, \ and\
  \bibinfo {author} {\bibfnamefont {D.}~\bibnamefont {Cho}},\ }\bibfield
  {title} {\enquote {\bibinfo {title} {{Optical Stern-Gerlach effect from the
  Zeeman-like ac Stark shift}},}\ }\href
  {http://dx.doi.org/10.1103/PhysRevA.65.033410} {\bibfield  {journal}
  {\bibinfo  {journal} {Phys. Rev. A}\ }\textbf {\bibinfo {volume} {65}},\
  \bibinfo {pages} {033410} (\bibinfo {year} {2002})}\BibitemShut {NoStop}%
\bibitem [{\citenamefont {Dehmelt}(1983)}]{Dehmelt:1983}%
  \BibitemOpen
  \bibfield  {author} {\bibinfo {author} {\bibfnamefont {H.}~\bibnamefont
  {Dehmelt}},\ }\bibfield  {title} {\enquote {\bibinfo {title} {{Stored-Ion
  Spectroscopy}},}\ }in\ \href {http://dx.doi.org/10.1007/978-1-4613-3715-7_6}
  {\emph {\bibinfo {booktitle} {Advances in Laser Spectroscopy}}}\ (\bibinfo
  {publisher} {Springer US},\ \bibinfo {address} {Boston, MA},\ \bibinfo {year}
  {1983})\ p.\ \bibinfo {pages} {153}\BibitemShut {NoStop}%
\bibitem [{\citenamefont {Dehmelt}(1986)}]{Dehmelt:1986}%
  \BibitemOpen
  \bibfield  {author} {\bibinfo {author} {\bibfnamefont {H.}~\bibnamefont
  {Dehmelt}},\ }\bibfield  {title} {\enquote {\bibinfo {title} {{Continuous
  Stern-Gerlach effect: Principle and idealized apparatus.}}}\ }\href
  {http://dx.doi.org/10.1073/pnas.83.8.2291} {\bibfield  {journal} {\bibinfo
  {journal} {Proc. Natl. Acad. Sci. U.S.A.}\ }\textbf {\bibinfo {volume}
  {83}},\ \bibinfo {pages} {2291} (\bibinfo {year} {1986})}\BibitemShut
  {NoStop}%
\bibitem [{\citenamefont {Alberti}\ \emph {et~al.}(2016)\citenamefont
  {Alberti}, \citenamefont {Robens}, \citenamefont {Alt}, \citenamefont
  {Brakhane}, \citenamefont {Karski}, \citenamefont {Reimann}, \citenamefont
  {Widera},\ and\ \citenamefont {Meschede}}]{Alberti:2016}%
  \BibitemOpen
  \bibfield  {author} {\bibinfo {author} {\bibfnamefont {A.}~\bibnamefont
  {Alberti}}, \bibinfo {author} {\bibfnamefont {C.}~\bibnamefont {Robens}},
  \bibinfo {author} {\bibfnamefont {W.}~\bibnamefont {Alt}}, \bibinfo {author}
  {\bibfnamefont {S.}~\bibnamefont {Brakhane}}, \bibinfo {author}
  {\bibfnamefont {M.}~\bibnamefont {Karski}}, \bibinfo {author} {\bibfnamefont
  {R.}~\bibnamefont {Reimann}}, \bibinfo {author} {\bibfnamefont
  {A.}~\bibnamefont {Widera}}, \ and\ \bibinfo {author} {\bibfnamefont
  {D.}~\bibnamefont {Meschede}},\ }\bibfield  {title} {\enquote {\bibinfo
  {title} {{Super-resolution microscopy of single atoms in optical
  lattices}},}\ }\href {http://dx.doi.org/10.1088/1367-2630/18/5/053010}
  {\bibfield  {journal} {\bibinfo  {journal} {New J. Phys.}\ }\textbf {\bibinfo
  {volume} {18}},\ \bibinfo {pages} {053010} (\bibinfo {year}
  {2016})}\BibitemShut {NoStop}%
\bibitem [{\citenamefont {Belmechri}\ \emph {et~al.}(2013)\citenamefont
  {Belmechri}, \citenamefont {F{\"o}rster}, \citenamefont {Alt}, \citenamefont
  {Widera}, \citenamefont {Meschede},\ and\ \citenamefont
  {Alberti}}]{Belmechri:2013}%
  \BibitemOpen
  \bibfield  {author} {\bibinfo {author} {\bibfnamefont {N.}~\bibnamefont
  {Belmechri}}, \bibinfo {author} {\bibfnamefont {L.}~\bibnamefont
  {F{\"o}rster}}, \bibinfo {author} {\bibfnamefont {W.}~\bibnamefont {Alt}},
  \bibinfo {author} {\bibfnamefont {A.}~\bibnamefont {Widera}}, \bibinfo
  {author} {\bibfnamefont {D.}~\bibnamefont {Meschede}}, \ and\ \bibinfo
  {author} {\bibfnamefont {A.}~\bibnamefont {Alberti}},\ }\bibfield  {title}
  {\enquote {\bibinfo {title} {{Microwave control of atomic motional states in
  a spin-dependent optical lattice}},}\ }\href
  {http://dx.doi.org/10.1088/0953-4075/46/10/104006} {\bibfield  {journal}
  {\bibinfo  {journal} {J. Phys. B: At. Mol. Phys.}\ }\textbf {\bibinfo
  {volume} {46}},\ \bibinfo {pages} {4006} (\bibinfo {year}
  {2013})}\BibitemShut {NoStop}%
\bibitem [{\citenamefont {Kuhr}\ \emph {et~al.}(2005)\citenamefont {Kuhr},
  \citenamefont {Alt}, \citenamefont {Schrader}, \citenamefont {Dotsenko},
  \citenamefont {Miroshnychenko}, \citenamefont {Rauschenbeutel},\ and\
  \citenamefont {Meschede}}]{Kuhr:2005}%
  \BibitemOpen
  \bibfield  {author} {\bibinfo {author} {\bibfnamefont {S.}~\bibnamefont
  {Kuhr}}, \bibinfo {author} {\bibfnamefont {W.}~\bibnamefont {Alt}}, \bibinfo
  {author} {\bibfnamefont {D.}~\bibnamefont {Schrader}}, \bibinfo {author}
  {\bibfnamefont {I.}~\bibnamefont {Dotsenko}}, \bibinfo {author}
  {\bibfnamefont {Y.}~\bibnamefont {Miroshnychenko}}, \bibinfo {author}
  {\bibfnamefont {A.}~\bibnamefont {Rauschenbeutel}}, \ and\ \bibinfo {author}
  {\bibfnamefont {D.}~\bibnamefont {Meschede}},\ }\bibfield  {title} {\enquote
  {\bibinfo {title} {{Analysis of dephasing mechanisms in a standing-wave
  dipole trap}},}\ }\href {http://dx.doi.org/10.1103/PhysRevA.72.023406}
  {\bibfield  {journal} {\bibinfo  {journal} {Phys. Rev. A}\ }\textbf {\bibinfo
  {volume} {72}},\ \bibinfo {pages} {023406} (\bibinfo {year}
  {2005})}\BibitemShut {NoStop}%
\bibitem [{\citenamefont {Alberti}\ \emph {et~al.}(2014)\citenamefont
  {Alberti}, \citenamefont {Alt}, \citenamefont {Werner},\ and\ \citenamefont
  {Meschede}}]{Alberti:2014}%
  \BibitemOpen
  \bibfield  {author} {\bibinfo {author} {\bibfnamefont {A.}~\bibnamefont
  {Alberti}}, \bibinfo {author} {\bibfnamefont {W.}~\bibnamefont {Alt}},
  \bibinfo {author} {\bibfnamefont {R.}~\bibnamefont {Werner}}, \ and\ \bibinfo
  {author} {\bibfnamefont {D.}~\bibnamefont {Meschede}},\ }\bibfield  {title}
  {\enquote {\bibinfo {title} {{Decoherence models for discrete-time quantum
  walks and their application to neutral atom experiments}},}\ }\href
  {http://dx.doi.org/10.1088/1367-2630/16/12/123052} {\bibfield  {journal}
  {\bibinfo  {journal} {New J. Phys.}\ }\textbf {\bibinfo {volume} {16}},\
  \bibinfo {pages} {123052} (\bibinfo {year} {2014})}\BibitemShut {NoStop}%
\bibitem [{\citenamefont {Kovachy}\ \emph {et~al.}(2015)\citenamefont
  {Kovachy}, \citenamefont {Asenbaum}, \citenamefont {Overstreet},
  \citenamefont {Donnelly}, \citenamefont {Dickerson}, \citenamefont
  {Sugarbaker}, \citenamefont {Hogan},\ and\ \citenamefont
  {Kasevich}}]{Kovachy:2015}%
  \BibitemOpen
  \bibfield  {author} {\bibinfo {author} {\bibfnamefont {T.}~\bibnamefont
  {Kovachy}}, \bibinfo {author} {\bibfnamefont {P.}~\bibnamefont {Asenbaum}},
  \bibinfo {author} {\bibfnamefont {C.}~\bibnamefont {Overstreet}}, \bibinfo
  {author} {\bibfnamefont {C.~A.}\ \bibnamefont {Donnelly}}, \bibinfo {author}
  {\bibfnamefont {S.~M.}\ \bibnamefont {Dickerson}}, \bibinfo {author}
  {\bibfnamefont {A.}~\bibnamefont {Sugarbaker}}, \bibinfo {author}
  {\bibfnamefont {J.~M.}\ \bibnamefont {Hogan}}, \ and\ \bibinfo {author}
  {\bibfnamefont {M.~A.}\ \bibnamefont {Kasevich}},\ }\bibfield  {title}
  {\enquote {\bibinfo {title} {{Quantum superposition at the half-metre
  scale}},}\ }\href {http://dx.doi.org/10.1038/nature16155} {\bibfield
  {journal} {\bibinfo  {journal} {Nature}\ }\textbf {\bibinfo {volume} {528}},\
  \bibinfo {pages} {530} (\bibinfo {year} {2015})}\BibitemShut {NoStop}%
\bibitem [{\citenamefont {Stamper-Kurn}\ \emph {et~al.}(2016)\citenamefont
  {Stamper-Kurn}, \citenamefont {Marti},\ and\ \citenamefont
  {M{\"u}ller}}]{StamperKurn:2016}%
  \BibitemOpen
  \bibfield  {author} {\bibinfo {author} {\bibfnamefont {D.~M.}\ \bibnamefont
  {Stamper-Kurn}}, \bibinfo {author} {\bibfnamefont {G.~E.}\ \bibnamefont
  {Marti}}, \ and\ \bibinfo {author} {\bibfnamefont {H.}~\bibnamefont
  {M{\"u}ller}},\ }\bibfield  {title} {\enquote {\bibinfo {title} {{Verifying
  quantum superpositions at metre scales}},}\ }\href
  {http://dx.doi.org/10.1038/nature19108} {\bibfield  {journal} {\bibinfo
  {journal} {Nature}\ }\textbf {\bibinfo {volume} {537}},\ \bibinfo {pages}
  {E1} (\bibinfo {year} {2016})}\BibitemShut {NoStop}%
\bibitem [{\citenamefont {Asb{\'o}th}\ \emph {et~al.}()\citenamefont
  {Asb{\'o}th}, \citenamefont {Rakovszk},\ and\ \citenamefont
  {Alberti}}]{Asboth:2016}%
  \BibitemOpen
  \bibfield  {author} {\bibinfo {author} {\bibfnamefont {J.}~\bibnamefont
  {Asb{\'o}th}}, \bibinfo {author} {\bibfnamefont {T.}~\bibnamefont
  {Rakovszk}}, \ and\ \bibinfo {author} {\bibfnamefont {A.}~\bibnamefont
  {Alberti}},\ }\href@noop {} {\enquote {\bibinfo {title} {Detecting bulk
  topological invariants of a periodically driven systems by dissipation},}\
  }\bibinfo {note} {Manuscript in preparation.}\BibitemShut {Stop}%
\end{thebibliography}%

\end{document}